# Aging and memory in a two-dimensional electron system in Si

J. Jaroszyński, Dragana Popović[*]

*National High Magnetic Field Laboratory, Florida State University, 1800 E. Paul Dirac Dr., Tallahassee, Florida 32310, USA*

**Abstract**

The relaxations of conductivity after a temporary change of carrier density $n_s$ during the waiting time $t_w$ have been studied in a strongly disordered two-dimensional electron system in Si. At low enough $n_s<n_g$ ($n_g$ – the glass transition density), the nonexponential relaxations exhibit aging and memory effects at low temperatures $T$. The aging properties change abruptly at the critical density for the metal-insulator transition $n_c<n_g$. The observed complex dynamics of the electronic transport is strikingly similar to that of other systems that are far from equilibrium.

## 1. Introduction

Low-density two-dimensional (2D) electron and hole systems in semiconductor heterostructures, where the existence of the metal and the metal-insulator transition (MIT) have been a subject of great interest and debate [1], represent particularly appealing model systems for studying the effects of the interplay of strong electronic correlations and disorder in a controlled and systematic way. While the competition between these two effects had been suggested a long time ago to lead to glassy dynamics [2], such electron or Coulomb glasses in general remain poorly understood. Recent theoretical proposals that describe the 2D MIT as the melting of a Coulomb glass [3,4] have found support in observations and studies of glassiness [5-9] in a 2D electron system (2DES) in Si. The onset of glassiness takes place at a density $n_g>n_c$ ($n_c$ – the critical density for the MIT), i.e. on the metallic side of the MIT, consistent with theory [4].

The resistance noise spectroscopy has revealed [5,6] a dramatic slowing down of the electron dynamics for $n_s<n_g$, accompanied by the onset of correlated statistics consistent with the hierarchical picture of glasses. Furthermore, studies of the nonexponential relaxations of conductivity σ [7], following a rapid change of $n_s$, show that the equilibration time $\tau_{eq}$ diverges exponentially as $T\rightarrow 0$, suggesting a glass transition at $T_g=0$. Here we report the results of a different relaxation protocol, which demonstrates that, for $n_s<n_g$, the 2DES exhibits aging and memory, the hallmarks of glassy dynamics.

## 2. Experiment

The measurements were performed on a 2DES in (100)-Si metal-oxide-semiconductor field-effect transistors with high disorder (the 4.2 K peak mobility ≈0.06 m$^2$/Vs at the applied substrate bias $V_{sub}= -2$ V). The device length × width were 1×90 μm$^2$ (sample A) and 2×50 μm$^2$ (sample B). Other sample details have been given elsewhere [5]. $n_s$ was varied by the gate voltage $V_g$, such that $n_s(10^{11}\text{cm}^{-2})$ =4.31($V_g$[V]-6.3); $n_g(10^{11}\text{cm}^{-2})$=(7.5±0.3) and $n_c(10^{11}\text{cm}^{-2})$=(4.5±0.4), where $n_g$ was determined from the onset of slow, correlated dynamics in noise and $n_c$ from $\sigma(n_s,T)$ measurements on both metallic and insulating sides [5,6] (see also Fig. 1). The error bars in $n_g$ and $n_c$ include the sample-to-sample variations. $\sigma$ was measured with the standard ac lock-in technique at ~13 Hz and 5−10 μV excitation voltage.

---

[*] Corresponding author. Tel.: +1-850-644-3913; fax: +1-850-644-5038; e-mail: dragana@magnet.fsu.edu.

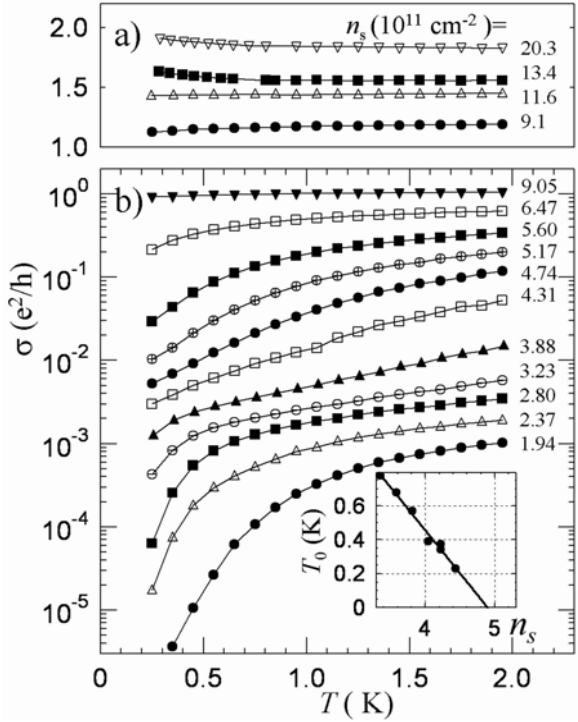

Figure 2 Sample A. $\sigma(T)$ for several $n_s$, as shown. Solid lines guide the eye. Inset: activation energies vs. $n_s$. The solid line is a linear fit. The vanishing of activation energies is often used as a criterion to determine $n_c$ [10].

Figure 1 demonstrates that, while $d\sigma/dT$=0 at $n_s^* \approx 12 \times 10^{11}\text{cm}^{-2}$, the vanishing of the activation energies in the insulating regime, where $\sigma \propto \exp(-T_0/T)$, yields $n_c$ that is much lower (Fig. 1 inset). The relaxation effects described below are observed only for $\sigma < 2$ $e^2/h$, i.e. $k_F l<1$ ($k_F$ – Fermi wave vector, $l$ – mean free path) [11].

The experimental protocol [Figs. 2(a) and 2(b)] starts with the 2DES, induced by an applied gate voltage $V_0$ (density $n_0$), in equilibrium at 10 K. The sample is then cooled to the measurement $T$, resulting in no visible relaxations on our experimental time scales, and the equilibrium $\sigma_0(n_0, T)$ is thus established. $V_g$ is then switched rapidly (within 1 s) to a different value $V_1$, where it is kept for a time $t_w$. Finally, it is changed back to $V_0$, and the slowly evolving $\sigma(t, n_0, T)$ is measured from the moment when $V_g$ reattains its original value $V_0$.

### 3. Results and discussion

Figure 2(c) shows some $\sigma(t)$ measured for a fixed $t_w$ at different $T$ for a given $V_0$ and $V_1$. Here we focus on the $t_w \ll \tau_{eq}(T)$ regime, where the system is not able to reach equilibrium under the new conditions during $t_w$. In the opposite case, when $\tau_{eq}(T) \ll t_w$ and the 2DES equilibrates at a new state corresponding to $V_1$, it has been established [8] that the relaxations do not depend on $t_w$, i.e. the system, naturally, has no memory of the waiting time.

However, when $t_w \ll \tau_{eq}(T)$ [corresponding to $T$< 3 K in Fig. 2(c)], $\sigma(t)$ clearly depend on history (*aging* effect), i.e. on $t_w$ during which the 2DES relaxes away from its initial equilibrium state determined by $V_0$ and towards a new equilibrium state determined by $V_1$. It can be also said that the system has a *memory* of the time it spent with $V_g=V_1$. We find that, for $n_0 < n_c$, the aging function $\sigma(t, t_w)$ can be collapsed onto a single curve simply by rescaling the time axis by $t_w$, as demonstrated in Figs. 3(a) and 3(b) for two different $T$. This is called simple or full aging. It has been observed also in other systems out-of-equilibrium, such as an electron glass in InO films deep in the insulating regime [12] and spin glasses [13].

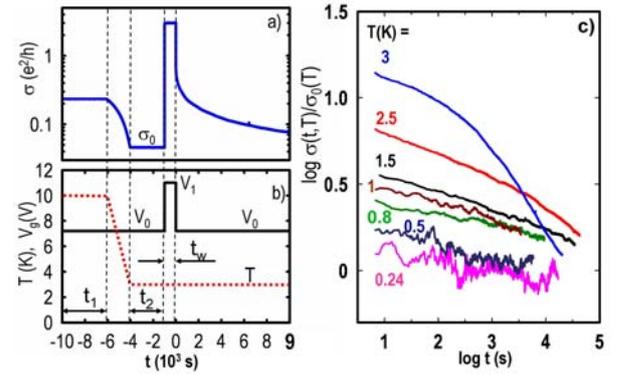

Figure 1 Sample A. (a) $\sigma(t)$ for $V_0$=7.2 V [$n_s(10^{11}\text{cm}^{-2})$ = 3.88 < $n_c$], $V_1$=11 V [$n_s(10^{11}\text{cm}^{-2})$ = 20.26], $t_w$=1000 s, and $T$=3.0 K. (b) The corresponding experimental protocol: $V_g(t)$ and $T(t)$. (c) The same as in (a), but for several $T$, as shown.

As soon as $n_s > n_c$, however, we find systematic deviations from full aging. Similar to other glassy materials [14,15], the data can be scaled then with a modified waiting time $t_w^\mu$ [Figs. 3(c) and 3(d)], where $\mu$ is a fitting parameter ($\mu=1$ for full aging) that, in this system, does not depend on $T$ [Figs. 3(c) and 3(d)]. Even though $\mu$ may not have a clear physical meaning, we adopt the $\mu$-scaling approach as a convenient measure of the departure from full aging.

Figure 3(e) shows our central result, $\mu$ as a function of $n_0$. There is a clear distinction between the full aging regime for $n_s < n_c$ (insulating phase), and the aging regime where significant departures from full scaling are seen. It is striking that the largest departure occurs at $n_s \approx n_g$. For $n_s > n_g$, it appears as if the full aging is restored, but that may be an artifact of trying to collapse very small (a few per cent) relaxations, gradually vanishing with $n_0$, accompanied by instrumental (white) noise of comparable magnitude [9].

Departures from full aging have been observed also in InO films, as a result of the increasing step size $\Delta V_g = V_1 - V_0$ [12]. That effect, however, cannot explain the departure from full aging that we observe for $n_c < n_s \leq n_g$. We have, indeed, verified experimentally that $\mu$ depends only on $n_0$ and not on $n_1$ or $\Delta V_g$ (as long as $t_w \ll \tau_{eq}(T)$ at $V_1$, of course, which is a prerequisite for aging). It is also easy to see from Fig. 3 that the size of $\Delta V_g$ is not relevant for $\mu(n_0)$. Namely, the data in Fig. 3(e) were obtained by increasing $n_0$ and keeping $n_1$ fixed, so that $\Delta V_g$ is smaller for higher $n_0$. The effect seen in InO films [12] would thus lead to better $t/t_w$ scaling at higher $n_0$, which is just the opposite of what we observe. It was also reported [12] that the same $\Delta V_g$ had a more detrimental effect on $t/t_w$ scaling in samples with lower density. Here, however, we observe just the opposite, i.e. that $t/t_w$ scaling works better at lower densities in spite of a larger $\Delta V_g$. We are forced to conclude that, even if the effect reported in [12] might be present also in our system, it is

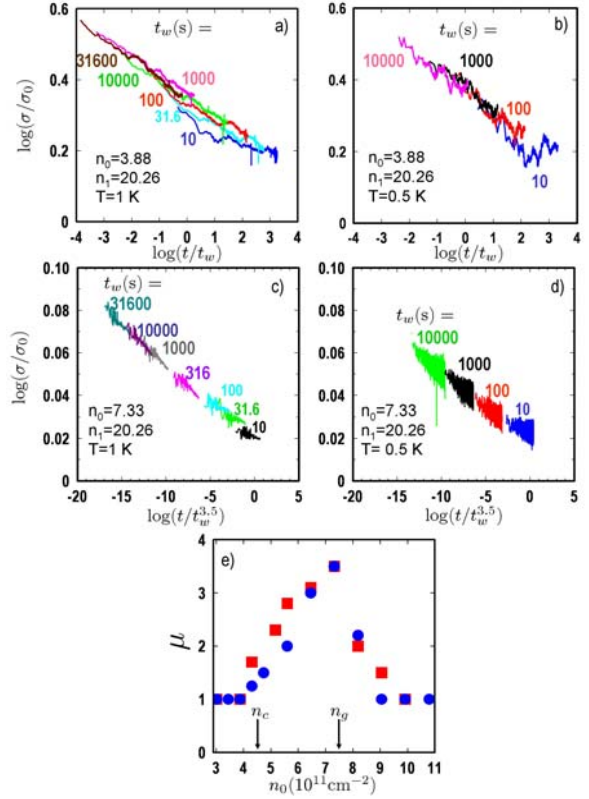

Figure 3 (a)-(d) Sample B. $\sigma(t)$ for two different $n_0(10^{11}\text{cm}^{-2})$ [(a), (b) and (c), (d), respectively] and a fixed $n_1(10^{11}\text{cm}^{-2})$ at $T=1$ K (a), (c) and $T=0.5$ K (b), (d). The data in (a), (b) are scaled with the waiting time $t_w$. For higher $n_0$, scaling with $t_w^\mu$ improves the collapse of the data (c), (d). (e) $\mu$ vs. $n_0$ (dots: sample A, squares: sample B). $\mu$ does not depend on $n_1$.

overshadowed by another, more dominant effect that leads instead to the departure from scaling at high $n_s$. That effect is the delocalization of the 2DES at the MIT. By contrast, electrons in InO films [12] always remain deep in the insulating phase so that other, weaker effects may become manifest.

If all the electrons are removed from the 2DES layer during $t_w$, it has been shown [8] that subsequent relaxations do not depend on $t_w$. This provides strong support for the existence of glassiness in the 2DES itself, instead of the slow, glassy dynamics being due to the response of the electrons to extrinsic slow degrees of freedom (e.g. a glassy background potential). In fact, even if the background disorder potential is glassy, then the abrupt change in the aging properties of the glassy phase precisely at the 2D MIT [Fig. 3(e)] indicates that the 2DES affects the properties of the background. In other words, this means that the 2D electrons cannot be thought of as simply following the time-dependent changes of the background that is independent of the 2DES, but rather that together they represent a coupled, strongly interacting system.

## 4. Conclusions

A detailed study of aging in a 2DES in Si shows an abrupt change in the nature of the glassy phase at the 2D MIT before it vanishes entirely at a higher $n_s$. While full aging is observed in the insulating phase ($n_s < n_c$), significant departures from full aging are seen in the intermediate, metallic glassy phase ($n_c < n_s \leq n_g$). We note that the mean-field models of glasses, for example, distinguish two different cases: one, where full aging is expected, and the other, where no $t/t_w$ scaling is expected [16]. Our results, therefore, put constraints on the theories of glassy freezing and its role in the physics of the 2D MIT.

We are grateful to I. Raičević for technical help and V. Dobrosavljević for discussions. This work was supported by NSF No. DMR-0403491, and NHMFL via NSF No. DMR-0654118. D.P. also acknowledges the hospitality of the Aspen Center for Physics, where this manuscript was written up.